\documentclass[english,prl,aps,showpacs,twocolumn,10.5pt]{revtex4-1}

\def\be{\begin{equation}}
\def\ee{\end{equation}}
\usepackage{graphicx}
\usepackage{bm}
\usepackage[T1]{fontenc}
\usepackage[latin9]{inputenc}
\usepackage{amsmath}
\usepackage{amssymb}
\usepackage{float}

\usepackage[bookmarks=true,
   colorlinks=true,
   linkcolor=blue,
   urlcolor=blue,
   citecolor=blue,
   bookmarks=true,
   hyperindex=true
]{hyperref}

\begin{document}
\title{Observation of quantum phase transition in spin-orbital-angular-momentum coupled Bose-Einstein condensate}

\author{Dongfang Zhang$^{1}$}
\email{These authors contribute equally to this work}

\author{Tianyou Gao$^{1,6}$}
\email{These authors contribute equally to this work}

\author{Peng Zou$^{2}$}
\email{These authors contribute equally to this work}

\author{Lingran Kong$^{1,6}$, Ruizong Li$^{1,6}$, Xing Shen$^{1,6}$, Xiao-Long Chen$^{4}$}

\author{Shi-Guo Peng$^{1}$}
\email{Electronic address: pengshiguo@wipm.ac.cn}

\author{Mingsheng Zhan$^{1,5}$, Han Pu$^{3,5}$}

\author{Kaijun Jiang$^{1,5}$}
\email{Electronic address: kjjiang@wipm.ac.cn}

\affiliation{$^{1}$State Key Laboratory of Magnetic Resonance and Atomic and Molecular Physics, Wuhan Institute of Physics and Mathematics, Chinese Academy of Sciences, Wuhan, 430071, China}
\affiliation{$^{2}$College of Physics, Qingdao University, Qingdao 266071, China}
\affiliation{$^{3}$Department of Physics and Astronomy, and Rice Quantum Institute, Rice University, Houston, Texas 77251, USA}
\affiliation{$^{4}$Centre for Quantum and Optical Science, Swinburne University of Technology, Melbourne 3122, Australia}
\affiliation{$^{5}$Center for Cold Atom Physics, Chinese Academy of Sciences, Wuhan, 430071, China}
\affiliation{$^{6}$School of Physics, University of Chinese Academy of Sciences, Beijing 100049, China}

\date{\today}

\begin{abstract}
  Orbital angular momentum (OAM) of light represents a fundamental optical freedom that can be exploited to manipulate quantum state of atoms. In particular, it can be used to realize spin-orbital-angular-momentum (SOAM) coupling in cold atoms by inducing an atomic Raman transition using two laser beams with differing OAM. Rich quantum phases are predicted to exist in many-body systems with SOAM coupling. Their observations in laboratory, however, are often hampered by the limited control of the system parameters. In this work we report, for the first time, the experimental observation of the ground-state quantum phase diagram of the SOAM coupled Bose-Einstein condensate (BEC). The discontinuous variation of the spin polarization as well as the vorticity of the atomic wave function across the phase boundaries provides clear evidence of first-order phase transitions. Our results open up a new way to the study of phase transitions and exotic quantum phases in quantum gases.
\end{abstract}

\maketitle

Coupling between a single particle's spin and orbital motion plays a crucial role in various many-body phenomena such as topological insulators and superconductors \cite{Kane2010RMPinsulator, Zhang2011RMPinsulator}. Spin-linear-momentum (SLM) coupling has been experimentally realized in ultracold Bose and Fermi gases \cite{Spielman2011NATUREsoc, Zhang2012PRLsoc, Zwierlein2012PRLsoc} and subsequently a variety of exotic quantum states have been predicted in theory and observed in experiment \cite{Dalibard2011RMPsoc, Spielman2013NATUREsoc, Zhai2015RPPsoc}. SLM coupling in cold atoms is achieved by inducing Raman transition in the atom with two counter-propagating laser fields. Recently several theoretical works proposed another fundamental type of spin-orbit coupling, namely the spin-orbital-angular-momentum (SOAM) coupling, and predicted rich quantum phases with first-order phase transitions in Bose condensate under such coupling \cite{Pu2015PRAsoam, Zhang2015PRAsoam, Zhang2015PRAsoamphase, Pu2016PRAsoam, Hu2015PRAsoam}. The SOAM coupling is achieved by inducing atomic Raman transition with a pair of co-propagating Laguerre-Gaussian (LG) laser fields that carries different orbital angular momenta. Both linear and angular momenta are important properties of quantum particles. The former possesses the spatial translational symmetry and has a continuous spectrum, whereas the latter takes the rotational symmetry and possesses a discrete spectrum. These fundamental differences render quantum gases subject to SOAM coupling unique for exploring exotic quantum transitions.

LG optical field contains a phase factor $e^{il\phi}$, where $\phi$ is the azimuthal angle and the integer $l$ is the winding number of the optical vortex \cite{Allen2003IOP}, and carries an orbital angular momentum (OAM) of $l \hbar$. Any coherent interaction of light and atoms needs to conserve energy, linear momentum and angular momentum. What happens to atoms if they encounter light carrying OAM is particularly intriguing as the OAM of atoms is quantized. When atoms interact with two LG beams with different OAMs, the relative winding phase of lights can be transferred to atoms in the transition between different spin states. Using this method, two LG beams have been used to diabatically write winding phases and spin textures into a freely expanding Bose-Einstein condensate (BEC), producing coreless vortices and Skyrmions in the process \cite{Bigelow2009PRLvortex, Bigelow2009PRLskymion}. The OAM of the LG beam has also been transferred to the metastable eigen state of ultracold atoms \cite{Lin2018arXivSOAM}. However, up until now, no experimental group has realized SOAM coupled ground state condensate, therefore the observation of the phase transition in the system is still elusive. Experimentally obtaining the ground state of the system is crucial in exploring phase transitions and providing quantitative comparisons with theoretical predictions.

Here we report, for the first time, the experimental observation of the ground-state quantum phase diagram of the SOAM coupled BEC. We adiabatically transfer the relative winding phase of two LG beams to the $^{87}$Rb Bose condensate, producing SOAM coupling of atoms in the transition between two spin states. The quantum phases are denoted by the order parameter of OAM number $l_{z}$ of atoms. We observe phase transition when the two-photon Raman coupling strength $\Omega_{R}$ and/or detuning $\delta$ approaches the critical value, by probing the spin-resolved coreless vortices. The phase diagram of the system in the $\Omega_{R}$-$\delta$ plane has been mapped out. The spin polarization across various phase boundaries exhibits discontinuous jumps, indictive of first-order phase transition.

\begin{figure}
\centerline{\includegraphics[width=8cm]{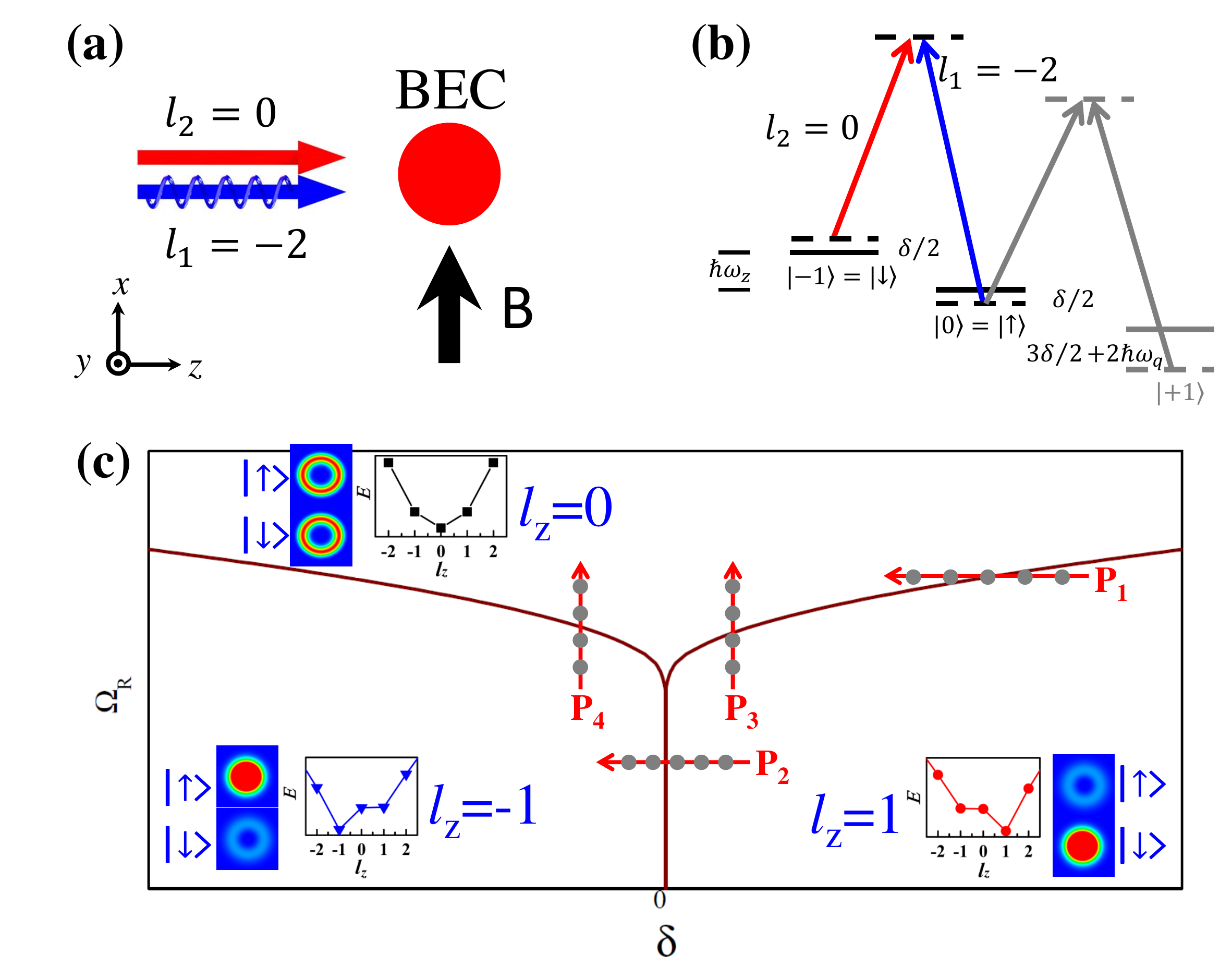}}
\caption{(color online) Scheme of the SOAM coupling. (a) Experimental schematics. Two laser beams with different orbital angular momentum ($l_{1}=-2$ and $l_{2}=0$) copropagate along the $z$ direction and interact with Rb BEC. The magnetic field is along the $x$ direction. (b) Level diagram. Two $\lambda = 790.02$ nm lasers couple two spin states $\left|\uparrow\right>=\left|F=1, m_{F}=0\right>$ and $\left|\downarrow\right>=\left|F=1, m_{F}=-1\right>$. $\delta$ is the two-photon detuning of the Raman lights. $\omega_{q}= 2\pi\times5.52$ kHz is the quadratic Zeeman shift. (c) The single-particle phase diagram. Three quantum phases are denoted by the quasi-orbital-angular momentum $l_{z}=1, 0, -1$, respectively. P$_{1}$ denotes the path crossing the phase transition from $l_{z}= 1$ to $l_{z}=0$ with a large Raman coupling strength $\Omega_{R}$. P$_{2}$, P$_{3}$ and P$_{4}$ denote other three paths crossing the phase transitions. The dots on the four paths denote the exemplary measurements in our experiment. The corresponding dispersion curve as well as the spin-dependent density distribution is also shown in each quantum phase.}  \label{Fig1}
\end{figure}

In the experiment, we produce a $^{87}$Rb BEC in a nearly spherical optical dipole trap with an atom number of $1.2(1)\times10^5$ as in our previous work \cite{Jiang2018Arxiv}. The mean trapping frequency $\bar{\omega}= \left(\omega_{x}+\omega_{y}+\omega_{z}\right)/3 = 2\pi\times 77.5$ Hz and the asphericity $A=(\omega_{max}-\omega_{min})/\bar{\omega}\approx3.7\%$, where $\omega_{max}$, $\omega_{min}$ are the maximum and minimum trapping frequencies along three directions, respectively. As shown in Fig. \ref{Fig1} (a) and (b), a pair of LG Raman beams ($l_{2}=0$ and $l_{1}=-2$) copropagate along the $z$ direction, suppressing the SLM coupling. The absolute winding number difference $\left|\Delta l\right|$ of the two LG beams equals to 2, which satisfies the requirement to observe the phase transition versus the variation of the Raman coupling strength \cite{Pu2015PRAsoam, Zhang2015PRAsoamphase}. The relative winding phase of the two light fields is transferred to the BEC in the Raman transition process between two atomic spin states $\left|\uparrow\right>=\left|F=1, m_{F}=0\right>$ and $\left|\downarrow\right>=\left|F=1, m_{F}=-1\right>$. A bias magnetic field produces a large quadratic Zeeman shift $\omega_{q}= 2\pi\times5.52$ kHz, which makes the effect of the spin state $\left|F=1, m_{F}=+1\right>$ negligibly small. We use the tune-out wavelength $\lambda = 790.02$ nm of the two LG beams \cite{Widera2016PRAtuneout}, in which the ground spin manifold of the Rb atom experiences no scalar AC stark shift. This implies that the vortex of BEC is produced due to the SOAM coupling, but not the trapping effect of the LG beams \cite{Bigelow2009PRLvortex}. We probe the spin-resolved spatial distributions of the two spin states in a time-of-flight (TOF) of 20 ms with the aid of a gradient magnetic field.

Our experimental scheme generates an effective single-particle Hamiltonian (see Supplementary Information for details)
\begin{equation}
\hat{H}_{0}=\hat{H}_{ho}+\frac{\delta}{2}\hat{\sigma}_{z}+
\Omega\left(\rho\right)\hat{\sigma}_{x}-\frac{l\hbar}{M\rho^{2}}\hat{L}_{z}\hat{\sigma}_{z}
+\frac{\left(l\hbar\right)^{2}}{2M\rho^{2}},\label{SPHam}
\end{equation}
 where $l=\left(l_{1}-l_{2}\right)/2$, $\hat{\sigma}_{x,z}$ are $2\times2$ Pauli matrices, $\hat{H}_{ho}\equiv\left(-\hbar^{2}/2M\right)\nabla^{2}+M\omega^{2}r^{2}/2$ with ${\bf{r}}=\left(\rho,\phi,z\right)$, $\omega$ is the trapping frequency, $\hat{L}_{z}=-i\hbar\partial_{\phi}$ denotes the quasi-orbital-angular-momentum operator of atoms along the $z$ axis, $\delta$ is the two-photon detuning, $\Omega\left(\rho\right)=\Omega_{R}\left(\rho/w\right)
^{\left|l_{1}\right|+\left|l_{2}\right|}e^{-2\rho^{2}/w^{2}}$ represents the spatial dependent Raman coupling characterized by the coupling strength $\Omega_R$, and $w$ is the waist of the two Raman beams. In our experiment the waists of the two Raman beams are almost the same with $w\approx63$ $\mu$m. Here the SOAM coupling is manifested as the term $\propto \hat{L}_{z}\hat{\sigma}_{z}$, which couples bare atomic states $\left|\uparrow, l_{\uparrow}=l_{z}-l\right>$ and $\left|\downarrow, l_{\downarrow}=l_{z}+l\right>$ with different orbital angular momentums $l_{\uparrow}$ and $l_{\downarrow}$ as measured in the lab frame.

Obviously, the quasi-orbital-angular momentum $l_z$ of a single atom is conserved, and then the ground-state quantum phases of the system in the absence of interactions can be characterized by definite $L_{z}$. The phase diagram of the single-particle Hamiltonian is shown in Fig. \ref{Fig1} (c). Due to the quantization of the OAM, the SOAM coupled BEC has three quantum phases denoted by the quasi-orbital-angular momentum of the ground state potential minimum, $l_{z}=1, 0, -1$ \cite{Pu2015PRAsoam}, respectively, which can be translated to the lab-frame orbital angular momentum as $\left(l_{\uparrow},l_{\downarrow}\right)=\left(2,0\right),\left(1,-1\right),\left(0,-2\right)$. The exemplary dispersion relation as well as the spin-dependent density distribution in each quantum phase is shown to clarify this claim. In the presence of interatomic interactions, it has been known that different ground-state quantum phases can still be distinguished by single-particle quasi-angular momentum $l_{z}$ except for the stripe phase, in which $l_{z}$ is no longer conserved \cite{Pu2015PRAsoam, Zhang2015PRAsoamphase}. Therefore, we may identify different quantum phases experimentally from the spatial structures of spin-up and -down atom clouds. The quantum phase $l_{z}=1$ corresponds to the Gaussian distribution in  $\left|\downarrow\right>$ and vortex structure in $\left|\uparrow\right>$, $l_{z}=-1$ corresponds to the vortex structure in  $\left|\downarrow\right>$ and Gaussian distribution in $\left|\uparrow\right>$, and $l_{z}=0$ corresponds to the vortex structures in both spin states.

\begin{figure}
\centerline{\includegraphics[width=8cm]{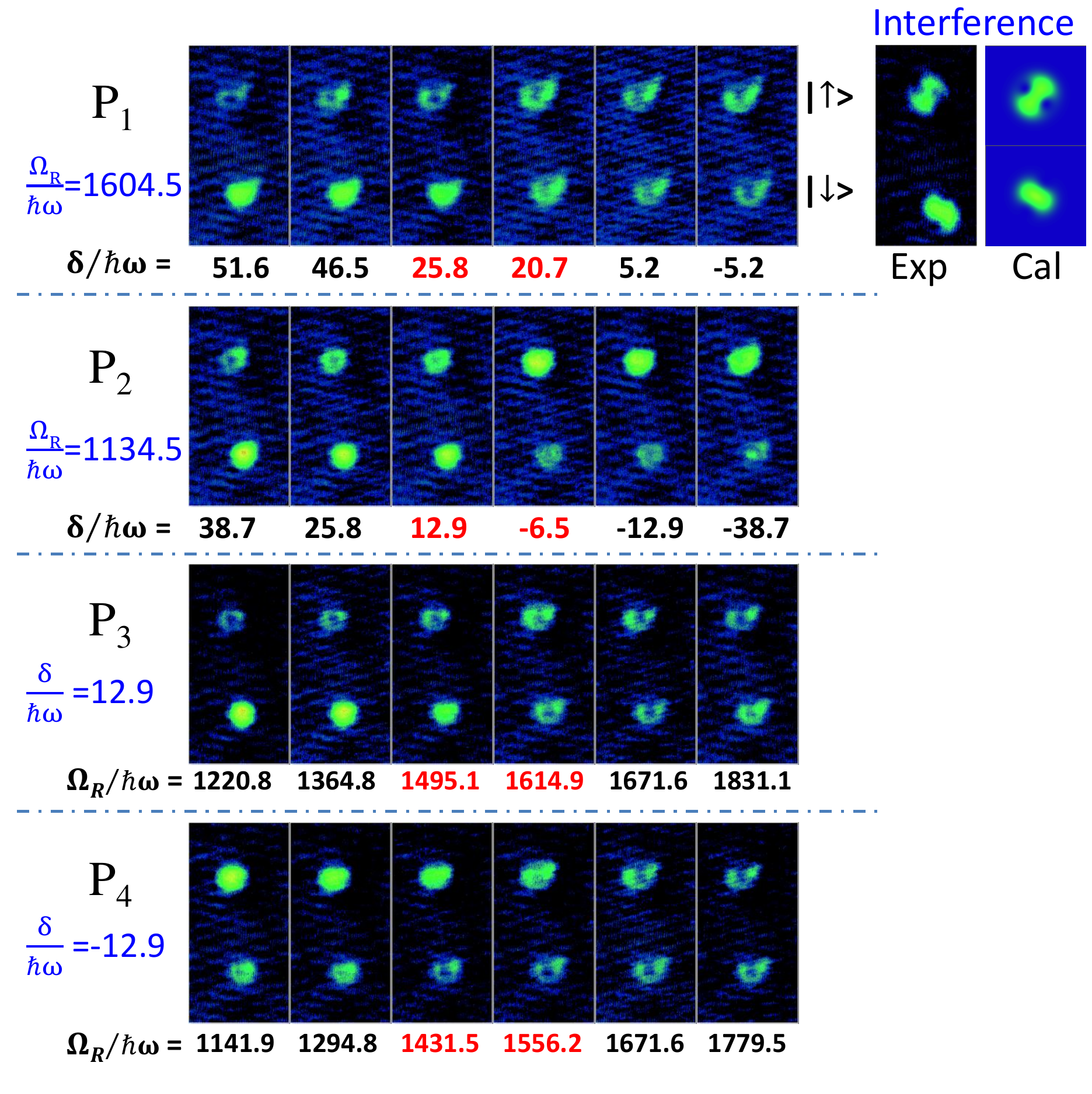}}
\caption{(color online) Observation of the phase transitions. P$_{1}$, P$_{2}$, P$_{3}$ and P$_{4}$ denote the four paths indicated in Fig. \ref{Fig1} (c). P$_{1}$ denotes the path with a large Raman coupling strength $\Omega_{R}/{\hbar\omega}=1604.5$. When the detuning $\delta$ decreases, the spin-dependent distributions change from Gaussian distribution in $\left|\downarrow\right>$ and vortex structure in $\left|\uparrow\right>$ to the vortex structures in two spin states, indicating the phase transition from $l_{z}=1$ to $l_{z}=0$. In the quantum phase $l_{z}=1$, two components $l=0$ and $l=2$ interfere in each of the two spin states with the aid of RF coupling, which are compared to the numerical calculations; Along the path P$_{2}$ with a small Raman coupling strength $\Omega_{R}/{\hbar\omega}=1134.5$, when $\delta$ decreases, the spin-dependent distributions change from Gaussian distribution in $\left|\downarrow\right>$ and vortex structure in $\left|\uparrow\right>$ to vortex structure in $\left|\downarrow\right>$ and Gaussian distribution in $\left|\uparrow\right>$, indicating the phase transition from $l_{z}=1$ to $l=-1$; Along the path P$_{3}$ with a positive detuning $\delta/{\hbar\omega} = 12.9$, when $\Omega_{R}$ increases, the spin-dependent distributions change from Gaussian distribution in $\left|\downarrow\right>$ and vortex structure in $\left|\uparrow\right>$ to vortex structures in two spin states, indicating the phase transition from $l_{z}=1$ to $l=0$; Along the path P$_{4}$ with a negative detuning $\delta/{\hbar\omega} = -12.9$, when $\Omega_{R}$ increases, the spin-dependent distributions change from vortex structure in $\left|\downarrow\right>$ and Gaussian distribution in $\left|\uparrow\right>$ to vortex structures in two spin states, indicating the phase transition from $l_{z}=-1$ to $l=0$. The uncertainties in determining the phase transition along the four paths are denoted by red numerical values under the atomic images.}  \label{Fig2}
\end{figure}

Four exemplary paths across various phase boundaries, denoted by P$_{1}$, P$_{2}$, P$_{3}$ and P$_{4}$, are shown in Fig. \ref{Fig1} (c). Along the paths P$_{1}$ and P$_{2}$, the Raman coupling strength $\Omega_{R}$ is fixed, while the detuning $\delta$ is varied. Along the paths P$_{3}$ and P$_{4}$, $\delta$ is fixed while $\Omega_R$ is varied.

To make the system adiabatically evolve in the ground state, we initially prepare atoms in the spin state $\left|\downarrow\right>$ by setting a large two-photon detuning $\delta_{i}/{2\pi\hbar}=400$ kHz. We then switch on the coupling strength $\Omega_R$ in 10 ms and then ramp the detuning $\delta$ adiabatically to the desired value in 150 ms. Fig. \ref{Fig2} shows the spin-resolved density distributions measured along the four paths. Along path P$_{1}$, we fix $\Omega_{R}/\hbar\omega=1604.5$. When $\delta$ is large positive, only the spin state $\left|\uparrow\right>$ has a vortex structure in the density distribution, while the spin state $\left|\downarrow\right>$ remains a Gaussian distribution. This region corresponds to the quantum phase $l_{z}=1$. We perform an interferometric measurement to detect OAMs of the two spin states $\left|\uparrow\right>$ and  $\left|\downarrow\right>$ \cite{Cornell1999PRLvortex, Phillips2006PRLvortex, Bigelow2009PRLvortex}. A resonant radio frequency (RF) pulse ($\tau\approx10 \mu$s) is shined on atoms before turning on the gradient magnetic field during the TOF, transferring a small fraction of atomic population back and forth between the two spin states. The two components $l=0$ and $l=2$ interfere in each spin state. In spin state $\left|\uparrow\right>$, the population of $l=0$ is comparable to that of $l=2$ and therefore the interference visibility is good. While the large population ratio between $l=0$ and $l=2$ in state $\left|\downarrow\right>$ makes the interference visibility weak. We show the numerical simulations of the interference patterns for the comparison. These measurements indicate that $l_{\downarrow}=0$ and $l_{\uparrow}=2$ in this quantum phase. When $\delta$ decreases, both spin states $\left|\downarrow\right>$ and $\left|\uparrow\right>$ exhibit vortex structures, indicating that the system has gone across the phase boundary to the quantum phase $l_{z}=0$.

\begin{figure}
\centerline{\includegraphics[width=8cm]{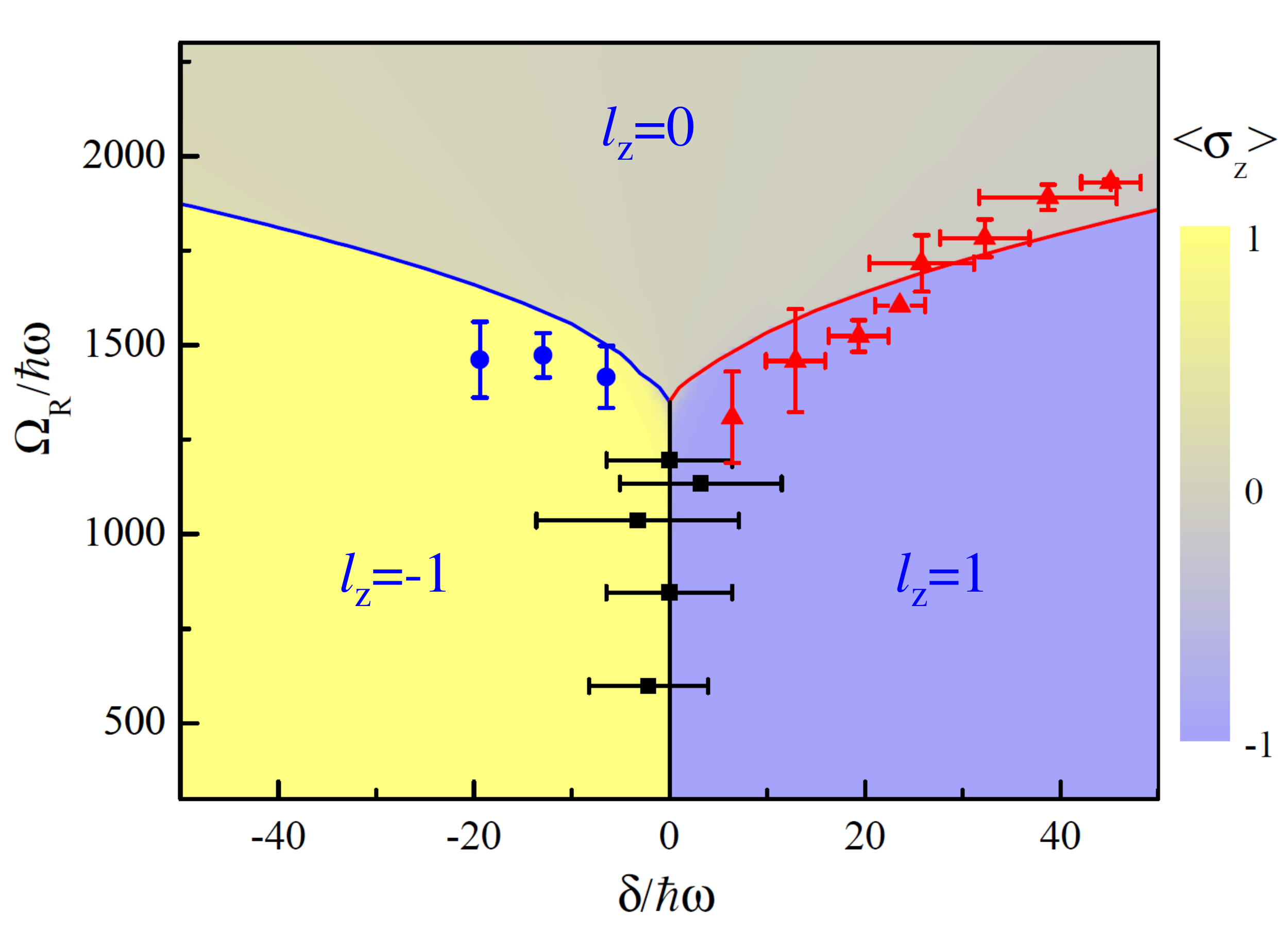}}
\caption{(Color online) Phase diagram of the SOAM coupled BEC. Three phases are denoted by the single-particle quasi-orbital-angular momentum $l_{z}=1, 0, -1$, respectively. The solid curves denote the calculated phase boundaries including the many-body interaction. The phase transitions are measured as in Fig. \ref{Fig2}. The error bars indicate the experimental uncertainties in determining the phase transitions. The error bars of the red points come from the measurements along the path P$_{1}$ and P$_{3}$, the error bars of the black points from the measurements along the path P$_{2}$, and the error bars of the blue points from the measurements along the path P$_{4}$. The colour denotes the spin polarization $\langle\sigma_{z}\rangle=(N_{\uparrow}-N_{\downarrow})/(N_{\uparrow}+N_{\downarrow})$.} \label{Fig3}
\end{figure}

We use the similar method to observe the phase transitions along the other three paths. P$_{2}$ path with a fixed Raman coupling strength $\Omega_{R}/{\hbar\omega}=1134.5$ and varying $\delta$ crosses the phase boundary between $l_{z}=1$ to $l=-1$, P$_{3}$ path with a fixed positive detuning $\delta/{\hbar\omega} = 12.9$ and varying $\Omega_R$ crosses the phase bounary between from $l_{z}=1$ to $l_{z}=0$, and finally P$_{4}$ path with a fixed negative detuning $\delta/{\hbar\omega} = -12.9$ and varying $\Omega_R$ crosses the phase boundary between $l_{z}=-1$ to $l_{z}=0$. In a very small region around the phase transition, the structure of each spin state fluctuates when repeating the measurements. These regions are also indicated in Fig. \ref{Fig2} and used to determine the experimental uncertainties of the phase transitions.

The observation along these and tens of other paths help us to map out the phase diagram which is presented in Fig. \ref{Fig3}. Numerical calculations including the atomic interaction (see the Supplementary Information) are also shown together. The phase diagram of the system is composed of three quantum phases denoted by $l_{z}=1, 0, -1$, respectively. Boundaries between these various phases are clearly distinguished both experimentally and theoretically. It is noted that the phase boundary including the many-body interaction has a small shift relative to that calculated from the single-particle Hamiltonian, which is smaller than our experimental uncertainty.

\begin{figure}
\centerline{\includegraphics[width=8cm]{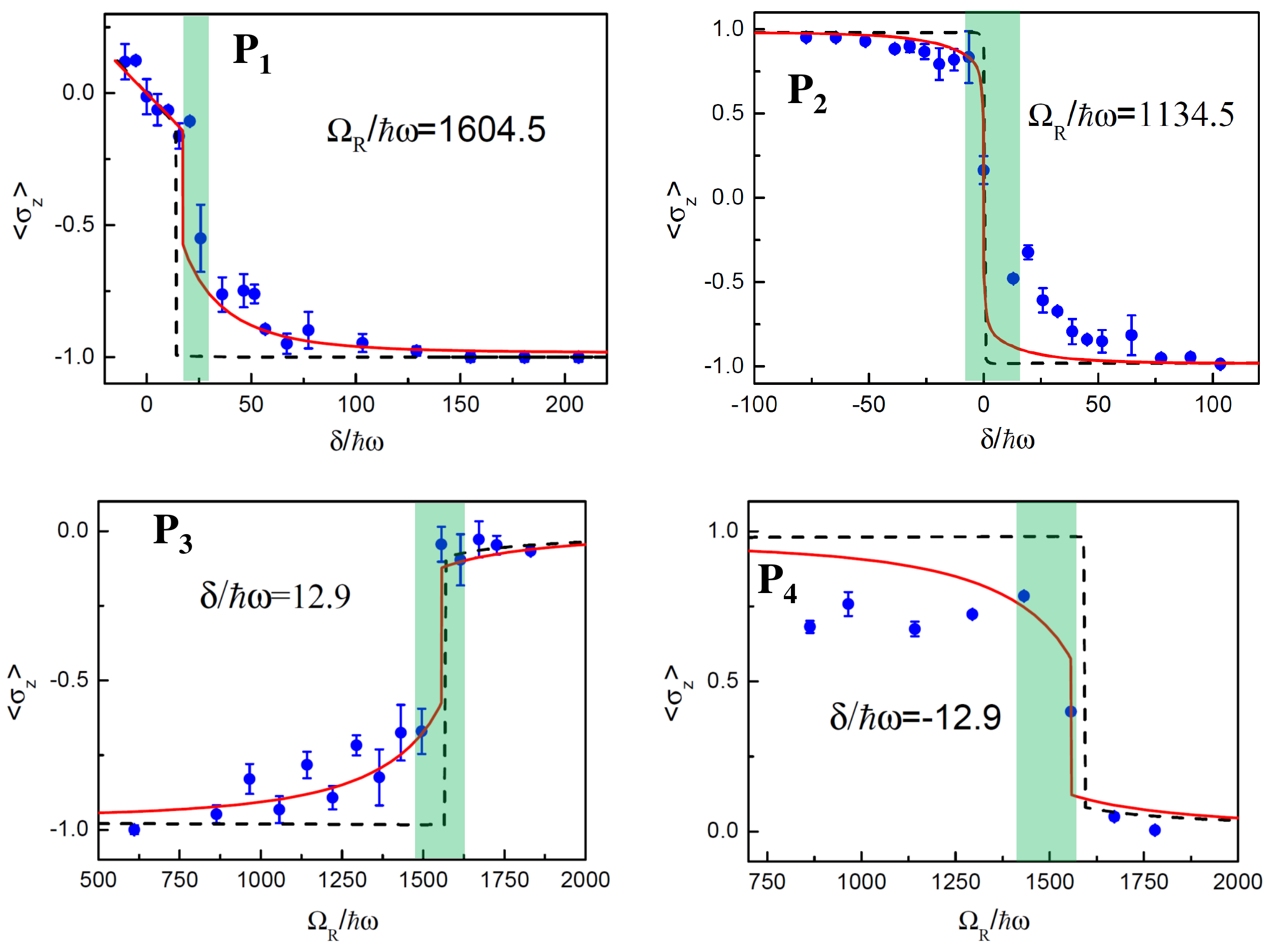}}
\caption{(Color online) Spin polarization $\langle\sigma_{z}\rangle=(N_{\uparrow}-N_{\downarrow})/(N_{\uparrow}+N_{\downarrow})$ across the phase transitions. P$_{1}$, P$_{2}$, P$_{3}$ and P$_{4}$ denote the four paths indicated in Fig. \ref{Fig1} (c). The black dashed curves are the calculations for $T=0$. The red solid curves are the calculations including finite-temperature effects. $T/T_{c}=0.32$ and $T_{c}$ is the BEC threshold temperature without SOAM coupling. The error bars mean the standard deviations of experimental measurements. The shaded areas indicate the phase transition regions determined from the measurements in Fig. \ref{Fig2}. }\label{Fig4}
\end{figure}

In Fig. \ref{Fig4}, we plot the spin polarization $\langle\sigma_{z}\rangle=(N_{\uparrow}-N_{\downarrow})/(N_{\uparrow}+N_{\downarrow})$ along the four paths. Our theoretical calculation for zero temperature indicate that $\langle\sigma_{z}\rangle=\mp1$ for the quantum phases $l_{z}=\pm1$, while $\langle\sigma_{z}\rangle\approx 0$ for the quantum phase $l_{z}=0$, i.e., the amplitude and sign of $\langle\sigma_{z}\rangle$ are locked by the value of $l_{z}$. This exotic behavior originates from the quantization of the OAM. Across the phase transition, $\langle\sigma_{z}\rangle$ jumps from one value to the other. The finite temperature affects the behavior of the spin polarization (see Supplementary Information). In our experiment ($T/T_{c}<0.4$, $T_{c}$ is the BEC threshold temperature without SOAM coupling), the condensate fraction is bigger than 90$\%$. The measured spin polarizations clearly show the jumping behavior across phase boundaries, which are in good quantitative agreements with the finite-temperature calculations. The phase transition regions determined by the discontinuous jump in spin polarization and by the spin-resolved density distribution almost completely overlap with each other. The discontinuous variation of the spin polarization provides the evidence of the first-order phase transition \cite{Zhang2015PRAsoamphase, Zhang2015PRAsoam}. It is noted that higher temperature will conceal the first-order phase transition and make the spin polarization variation smooth.

In summary, SOAM coupling is a new method in manipulating quantum states, which paves the way to explore exotic phase transitions. SOAM coupled BEC can be used to probe topological states like Half-skyrmion, vortex-antivortex, pairs and Mermin-Ho vortex and the meron pair \cite{Hu2015PRAsoam}. Richer quantum phases exist in the SOAM coupled BEC with higher order LG beams \cite{Zhang2015PRAsoam}. SOAM coupling can also be extended to systems with higher spins \cite{Pu2016PRAsoam} and Fermi gases. Our work represents a pioneering study on the SOAM coupling, which will stimulate further theoretical and experimental works to find novel quantum states.

Theoretically, it is predicted that there could exist a stripe phase in the ground state phase diagram, which represents roughly a superposition of states with different quasi-angular momenta \cite{Pu2016PRAsoam, Zhang2015PRAsoam, Zhang2015PRAsoamphase}. In the SLM coupling system, the corresponding stripe phase exhibits density oscillations on a spatial scale $\sim \lambda$ (where $\lambda$ is the optical wavelength), which prevents its direct detection due to the diffraction limit of the optical imaging systems. Here the stripe phase breaks the rotational symmetry and, with a spatial size on the order of the LG beam waist, should be readily detectable \cite{Pu2015PRAsoam,Pu2016PRAsoam, Zhang2015PRAsoamphase}. However, since the stripe phase only occupies a very small region in the phase diagram, our current experimental resolution is not sufficient to detect this phase. In the future, we plan to improve the experimental stability to observe the stripe phase. Alternatively, we will also consider the full spin-1 hyperfine ground manifold of the $^{87}$Rb, for which the stripe phase may occupy a larger parameter region in the phase diagram \cite{Pu2016PRAsoam}.

\subsection*{Acknowledgments}
We acknowledge fruitful discussions with Hui Hu, Xia-Ji Liu, Chunlei Qu and Hui Zhai. We also thank Hui Zhai for a critical reading of the manuscript. This work has been supported by the NKRDP (National Key Research and Development Program) under Grant No. 2016YFA0301503, NSFC (Grant No. 11474315, 11674358, 11434015, 11747059) and CAS under Grant No. YJKYYQ20170025. HP acknowledges support from the US NSF and the Welch Foundation (Grant No. C-1669).

\newpage
\begin{widetext}
\appendix

\section*{SUPPLEMENTARY INFORMATION}
\subsection*{Experimental setup and time sequence}

The experimental setup is shown in Fig. \ref{Figure6}(a). A pair of optical dipole beams form a spherical trap with the aid of the gravity as in our previous work \cite{Jiang2018ArxivSpp}. A pair of Helmholtz coils produce a bias magnetic field $B_{0}$, which provides the quantum axis and a large quadratic Zeeman shift $\omega_{q}= 2\pi\times5.52$ kHz of the Rb ground spin states. A pair of anti-Helmholtz coils produce a pulse of a gradient magnetic field $B^{'}$. A pair of Laguerre-Gaussian (LG) Raman beams ($l_{2}=0$ and $l_{1}=-2$) interact with the condensate, transferring the relative winding phase of two beams to atoms and producing spin-orbital-angular-momentum (SOAM) coupling. In fact, we initially try the experiment with a pair of LG Raman beams $l_{2}=1$ and $l_{1}=-1$, while in this case there are two technical challenges which makes the experiment difficult. Firstly, the size of BEC is about 10 $\mu$m, so it is required to make the relative position stability about 1 $\mu$m between any two of the LG beams and BEC during the long adiabatically ramping process in producing SOAM coupling (about 100 ms), maintaining two cores of the LG beams centered on the BEC. Secondly, the central optical intensities of the LG beams $l_{2}=1$ and $l_{1}=-1$ are both very weak, so we need high power of Raman lights to observe the phase transition. The choice of LG beams $l_{2}=0$ and $l_{1}=-2$ softens these two strict requirements and facilitates our experiment. The probe beam counterpropagates with the Raman beams, probing the density distribution of BEC in the $x-y$ plane.

\begin{figure}
\centerline{\includegraphics[width=15cm]{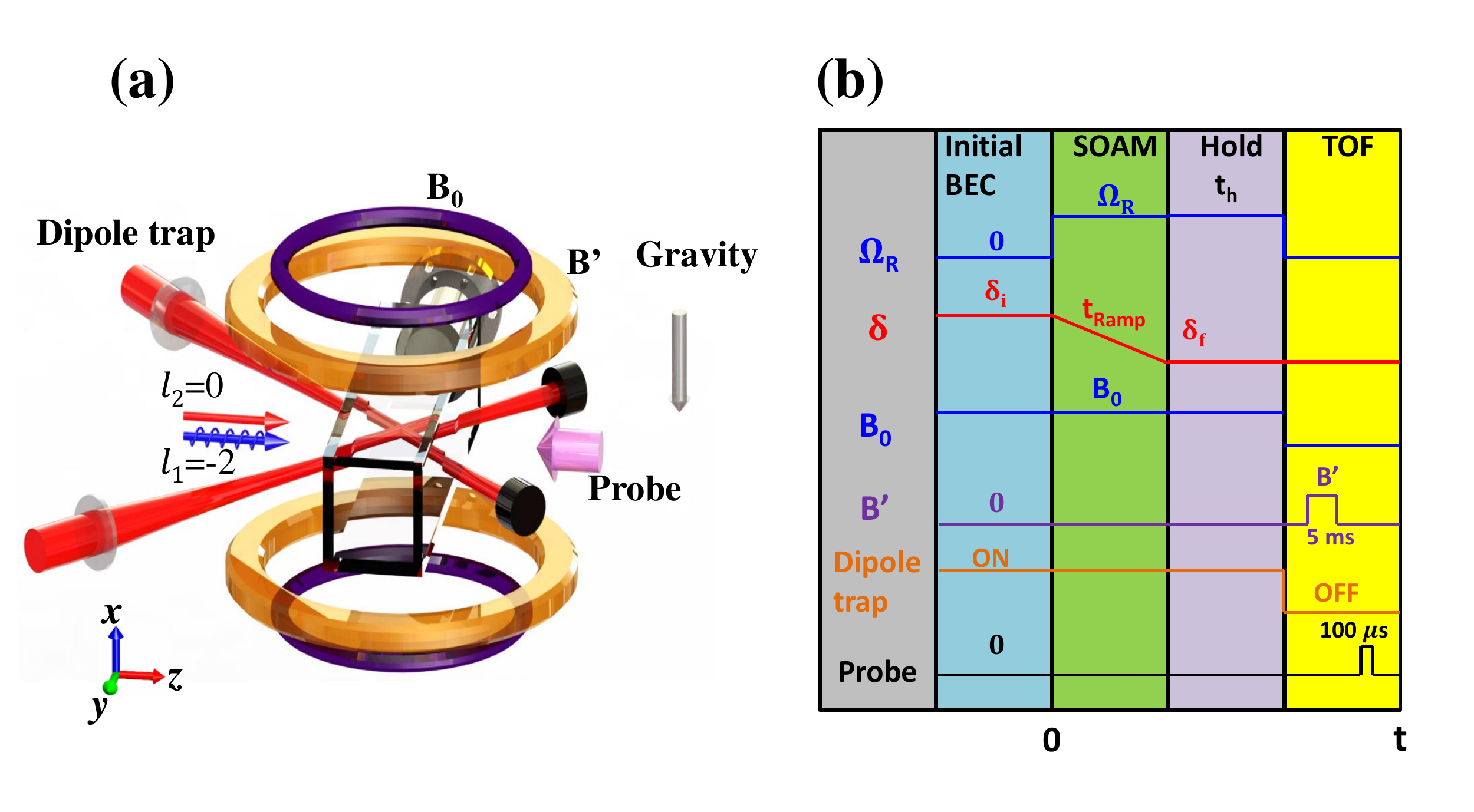}}
\caption{(color online) (a) Experimental setup. A pair of optical dipole beams orthogonally cross the vacuum chamber, producing Rb BEC. The gravity is along the $-x$ direction. A pair of Helmholtz coils produce a bias magnetic field $B_{0}$. A pair of anti-Helmholtz coils produce a gradient magnetic field $B^{'}$. A pair of LG Raman beams ($l_{2}=0$ and $l_{1}=-2$) copropagate along the z direction, interacting with the condensate. The probe beam counterpropagates with the Raman beams. (b) Experimental time sequence. The two-photon detuning is ramped from $\delta_{i}$ to $\delta_{f}$ during the SOAM coupling. The gradient magnetic field $B^{'}$ lasts 5 ms to spatially separate two spin states. The probe pulse is 100 $\mu$s. }  \label{Figure6}
\end{figure}

The experimental time sequence is shown in Fig. \ref{Figure6} (b). We set a large positive initial two-photon detuning $\delta_{i}/{\hbar\omega}>1000$. During the SOAM coupling with the Raman beams and the bias magnetic field on, the two-photon detuning is ramped from $\delta_{i}$ to $\delta_{f}$ in 150 ms. The system is then hold about 20 ms for the equilibrium. The cold atoms freely expand 3 ms after suddenly switching off the optical dipole trap and then the bias field $B^{'}$ is turned on for 5 ms. In this case, the two spin states $\left|\uparrow\right>=\left|F=1, m_{F}=0\right>$ and $\left|\downarrow\right>=\left|F=1 , m_{F}=-1\right>$ are spatially separated due to the Stern-Gerlach effect. We probe cold atoms with a total time-of-flight (TOF) of 20 ms.

\subsection*{Demonstration of the adiabaticity in the SOAM coupling process}

\begin{figure}
\centerline{\includegraphics[width=15cm]{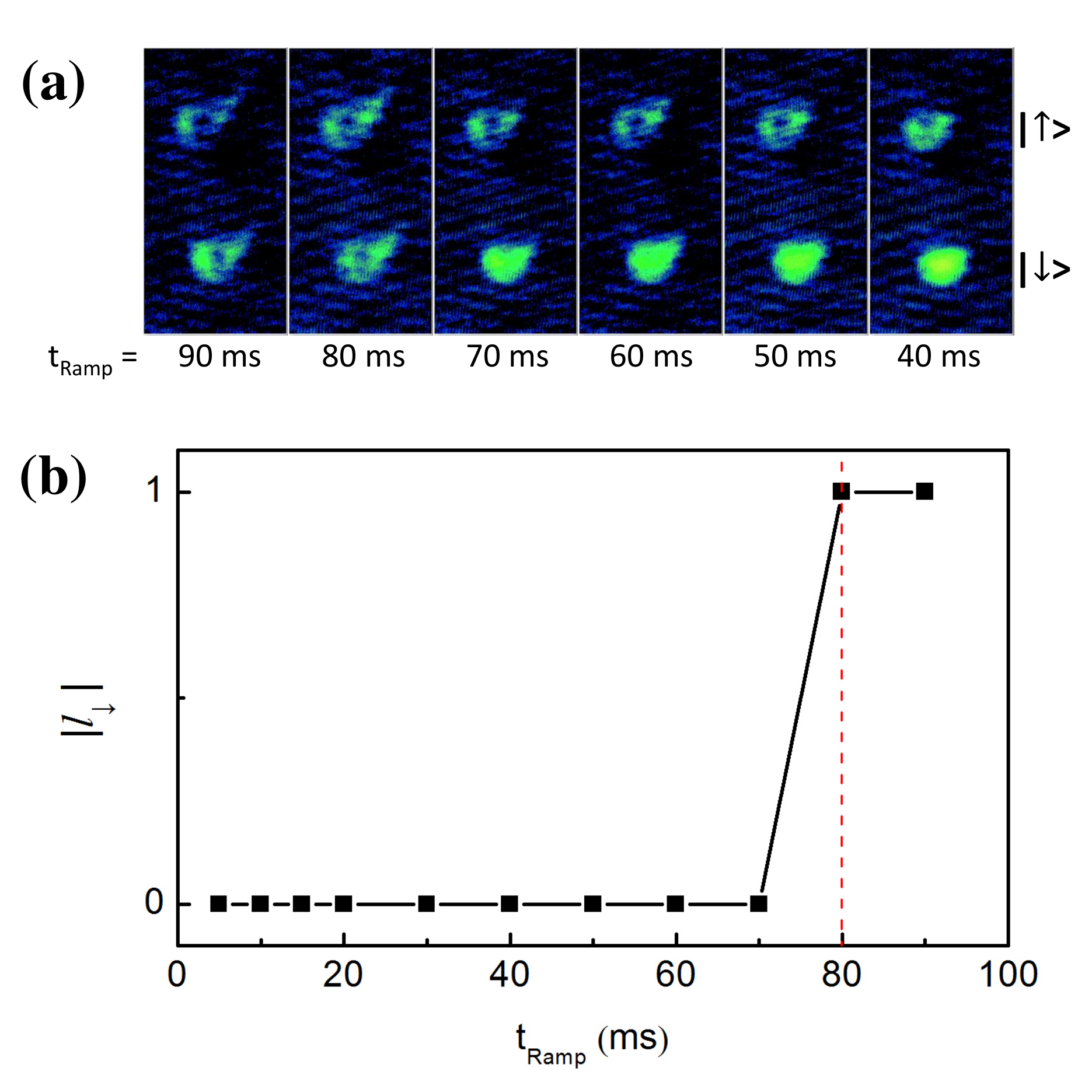}}
\caption{(Color online) Adiabaticity of the SOAM process along P$_{1}$ with $\Omega_{R}/{\hbar\omega}=1891.0$. The final detuning $\delta/{\hbar\omega} = 29.7$. The spin-dependent spatial distributions versus the ramping time $t_{Ramp}$ are shown in (a). The OAM number $l_{\downarrow}$ of the spin down state versus $t_{Ramp}$ is shown in (b). The dashed red line indicates the threshold time 80 ms for the adiabaticity. }\label{Fig5}
\end{figure}

A key to our work is to keep the system in the ground state. This is achieved by creating a ground state condensate without the Raman beams first and then turn on SOAM coupling adiabatically. We demonstrate the adiabaticity of the SOAM coupling process in our experiment. Here the two-photon detuning is ramped from the initial value to $\delta/{\hbar\omega} = 29.7$ with a constant Raman coupling strength $\Omega_{R}/{\hbar\omega}=1891.0$. As one can see from the ground-state phase diagram in Fig.~3 in the main text, if the ramping process is adiabatic, the system should stay in the quantum phase $l_{z}=0$ where both spin states have vortex structures. We change the ramping time and probe the spin-dependent density distributions in Fig. \ref{Fig5}. When the ramping time is less than 80 ms, only the spin state $\left|\uparrow\right>$ has the vortex distribution. For ramping times longer than 80 ms, both spin states have the vortex distributions. We can therefore infer that the threshold time for the adiabaticity is about 80 ms. In our experiment of mapping the phase diagram, the ramping time is 150 ms, which guarantees adiabaticity.

\subsection*{Calibrating the magnetic field using the RF adiabatic passage}

The absolute value and stability of the two-photon detuning $\delta$ is mainly determined by the bias magnetic field $B_{0}$. We calibrate the magnetic field by adiabatically coupling the ground spin states of $F=1$ with a radio frequency (RF) passage. In Fig. \ref{Figure7}, we scan the RF signal from 6.090 MHz to different values in 50 ms and simultaneously record the populations of the three spin states. The Hamitonian of the system dressed by the RF signal is

\begin{equation}
H=
\left (
  \begin{array}{ccc}
 \delta_{RF} & \Omega_{RF}/2  & 0   \\
  \Omega_{RF}/2  & \epsilon & \Omega_{RF}/2    \\
 0 &  \Omega_{RF}/2 & -\delta_{RF}      \\
  \end{array}
\right ), \label{eq:RFcoupling}
\end{equation}
where $\delta_{RF}/h = \nu_{RF} - \nu_{0}$ is the RF detuning. $h\nu_{0}= (E_{m_{F}=-1}-E_{m_{F}=1})/2$ is the effective resonant position, which is set by the bias magnetic field $B_{0}$. $\Omega_{RF}$ is the coupling strength of the RF signal. $\epsilon=E_{m_{F}=0}-(E_{m_{F}=-1}+E_{m_{F}=1})/2$ is the quadratic Zeeman shift. Then we can numerically calculate relative populations of the three spin states. From the comparison between the experimental measurements and the numerical calculations, we can deduce the bias magnetic field $B_{0}=8.807$ G. By repeating the measurements many times and observing the fluctuation of the resonance position, we can determine the magnetic field stability $\Delta B\approx 1$ mG.

\begin{figure}
\centerline{\includegraphics[width=15cm]{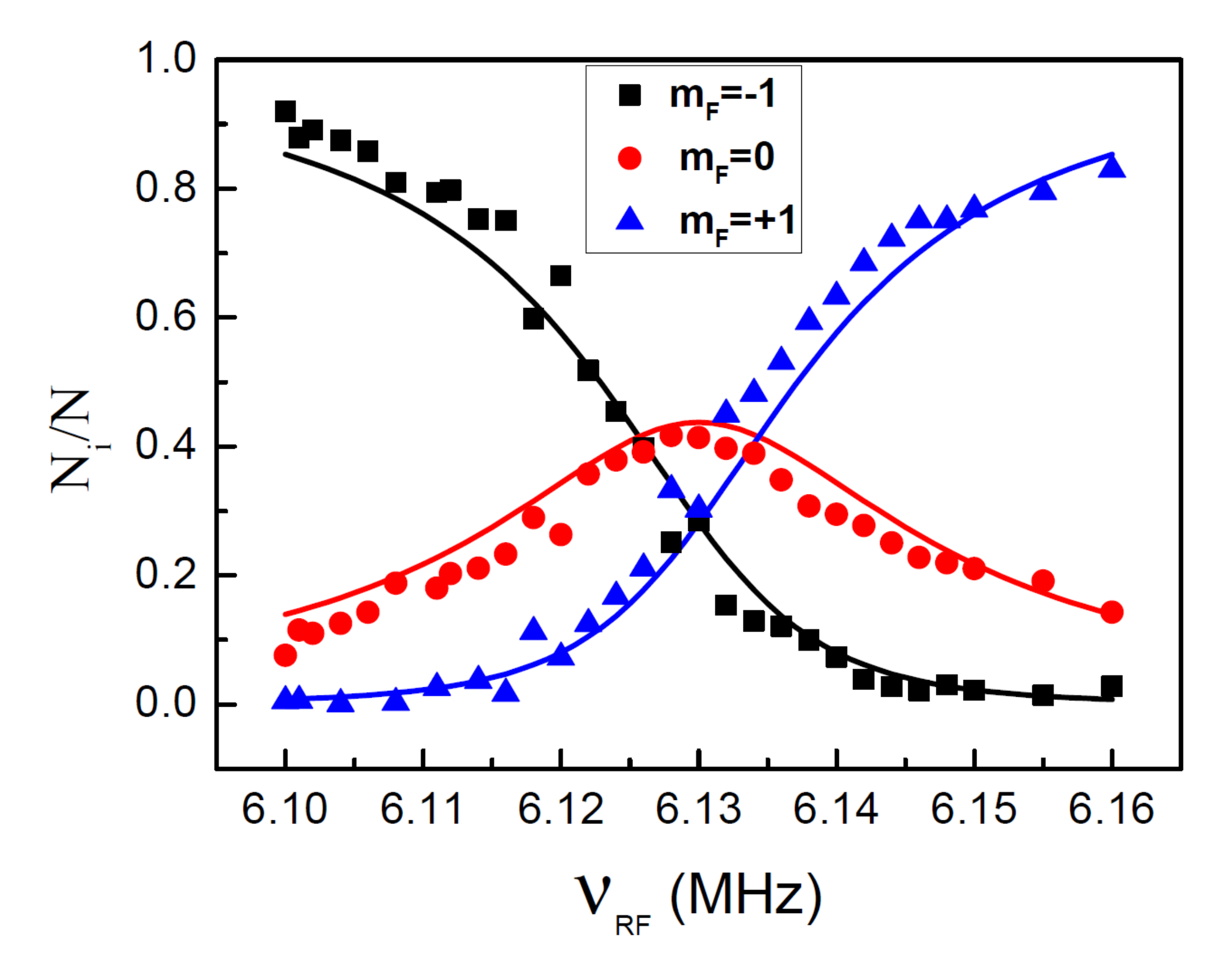}}
\caption{(color online) The spin population $N_{i}/N$ ($i=-1, 0, +1$) versus the radio frequency $\nu_{RF}$. $N_{i}/N$ ($i=-1, 0, +1$) denotes the population of spin state $\left|F=1, m_{F}\right>$ ($m_{F}=-1, 0, +1$), respectively. $N=\sum N_{i}$ is the total atom number. The solid curves are the numerical calculations of spin states.  }  \label{Figure7}
\end{figure}

\subsection*{Effective single-particle Hamiltonian}

In our experiment, a pair of Raman beams are applied along the $z$
axis with the amplitude
\begin{equation}
\mathcal{E}_{j}\left({\bf r}\right)=\sqrt{2I_{j0}}e^{-il_{j}\phi}\left(\frac{\rho}{w}\right)^{\left|l_{j}\right|}e^{-\rho^{2}/w^{2}}e^{ikz},\,\left(j=1,2\right)
\end{equation}
where ${\bf r}=\left(\rho,\phi,z\right)$ is the cylindrical coordinate,
$w$ is the width of the beam, and $I_{j0}$ is the intensity of the
$j$th beam. The phase winding $e^{-il_{j}\phi}$ reflects the orbital
angular momentum $-l_{j}\hbar$ carried by the beams. Then the intensity
profile of the Raman beams takes the form of
\begin{equation}
I_{j}\left(\rho\right)=\frac{1}{2}\left|\mathcal{E}_{j}\left({\bf r}\right)\right|^{2}=I_{j0}\left(\frac{\rho}{w}\right)^{2\left|l_{j}\right|}e^{-2\rho^{2}/w^{2}}.
\end{equation}
A bias magnetic field with strength $8.807$ G along the $x$ axis gives
a linear Zeeman splitting $\omega_{Z}=2\pi\times6.129$ MHz as well as
a quadratic Zeeman splitting $\omega_{q}=2\pi\times5.52$ kHz. Then the
Raman beams couples $\left|F=1,m_{F}=-1\right\rangle $ ($\left|\downarrow\right\rangle $)
and $\left|F=1,m_{F}=0\right\rangle $ states ($\left|\uparrow\right\rangle $),
while $\left|F=1,m_{F}=1\right\rangle $ state is far away from the
resonance. Therefore, our setup configuration is equivalent to a spin-half
system, and can be described by the following single-particle Hamiltonian,
\begin{equation}
\hat{\mathcal{H}}_{0}=-\frac{\hbar^{2}}{2M}\nabla^{2}+V_{T}\left({\bf r}\right)+V_{LG}\left(\boldsymbol{\rho}\right)
\end{equation}
with the Planck's constant $\hbar$ and atomic mass $M$, where $V_{T}\left({\bf r}\right)=M\omega^{2}r^{2}/2$
is the trapping potential with the frequency $\omega$, $V_{LG}\left(\boldsymbol{\rho}\right)$
denotes the interaction between the Laguerre-Gaussian (LG) beams and atoms in the $x-y$
plane and can be represented in the spin basis $\left\{ \left|\uparrow\right\rangle ,\left|\downarrow\right\rangle \right\} $
as \cite{Pu2015PRAsoamSpp}
\begin{equation}
V_{LG}\left(\boldsymbol{\rho}\right)=\left[\begin{array}{cc}
\delta/2 & 0\\
0 & -\delta/2
\end{array}\right]+\Omega\left(\rho\right)\left[\begin{array}{cc}
0 & e^{-i\left(l_{1}-l_{2}\right)\phi}\\
e^{i\left(l_{1}-l_{2}\right)\phi} & 0
\end{array}\right]
\end{equation}
 with the two-photon detuning $\delta$, and
\begin{equation}
\Omega\left(\rho\right)=\Omega_{R}\left(\frac{\rho}{w}\right)^{\left|l_{1}\right|+\left|l_{2}\right|}e^{-2\rho^{2}/w^{2}}
\end{equation}
describes the Raman coupling with the strength characterized by $\Omega_{R}\propto\left(I_{10}I_{20}\right)^{1/2}$.

In order to eliminate the $\phi$-dependence of $V_{LG}\left(\boldsymbol{\rho}\right)$,
we may introduce the following transformation to the single-particle
wave function $\tilde{\Psi}\left({\bf r}\right)=\left[\tilde{\psi}_{\uparrow}\left({\bf r}\right),\tilde{\psi}_{\downarrow}\left({\bf r}\right)\right]^{T}$,
\begin{equation}
\left[\begin{array}{c}
\tilde{\psi}_{\uparrow}\left({\bf r}\right)\\
\tilde{\psi}_{\downarrow}\left({\bf r}\right)
\end{array}\right]=\left[\begin{array}{cc}
e^{-il\phi} & 0\\
0 & e^{il\phi}
\end{array}\right]\left[\begin{array}{c}
\psi_{\uparrow}\left({\bf r}\right)\\
\psi_{\downarrow}\left({\bf r}\right)
\end{array}\right],
\end{equation}
where $l=\left(l_{1}-l_{2}\right)/2$. Then the single-particle Schr\"{o}dinger
equation $\hat{\mathcal{H}}_{0}\tilde{\Psi}\left({\bf r}\right)=E\tilde{\Psi}\left({\bf r}\right)$
becomes equivalent to $\hat{H}_{0}\Psi\left({\bf r}\right)=E\Psi\left({\bf r}\right)$.
Here, $\Psi\left({\bf r}\right)=\left[\psi_{\uparrow}\left({\bf r}\right),\psi_{\downarrow}\left({\bf r}\right)\right]^{T}$,
and the effective single-particle Hamiltonian $\hat{H}_{0}$ takes
the form of
\begin{equation}
\hat{H}_{0}=\left[\begin{array}{cc}
\hat{H}_{r}+\left(\hat{L}_{z}-l\hbar\right)^{2}/2M\rho^{2}+\delta/2 & \Omega\left(\rho\right)\\
\Omega\left(\rho\right) & \hat{H}_{r}+\left(\hat{L}_{z}+l\hbar\right)^{2}/2M\rho^{2}-\delta/2
\end{array}\right],
\end{equation}
where $\hat{H}_{r}=\left(-\hbar^{2}/2M\right)\left(\rho^{-1}\partial_{\rho}\rho\partial_{\rho}+\partial_{z}^{2}\right)+M\omega^{2}r^{2}/2$,
and $\hat{L}_{z}=-i\hbar\partial_{\phi}$ is the orbital angular momentum
operator along the $z$ axis. Alternatively, it can be written as
a very compact form as Eq.($1$) in the main text by using Pauli matrices.

\subsection*{Weakly interacting condensate and Gross-Pitaevskii equation}

When weak interatomic interactions are considered, an interacting
Bose-Einstein condensate is described by the Gross-Pitaevskii (GP)
equation in the mean-filed frame as $\left(\hat{H}_{0}+\hat{G}\right)\Psi\left({\bf r}\right)=\mu\Psi\left({\bf r}\right)$,
where $\mu$ is the chemical potential,
\begin{equation}
\hat{G}=\left[\begin{array}{cc}
g_{\uparrow\uparrow}\left|\psi_{\uparrow}\left({\bf r}\right)\right|^{2}+g_{\uparrow\downarrow}\left|\psi_{\downarrow}\left({\bf r}\right)\right|^{2} & 0\\
0 & g_{\downarrow\downarrow}\left|\psi_{\downarrow}\left({\bf r}\right)\right|^{2}+g_{\uparrow\downarrow}\left|\psi_{\uparrow}\left({\bf r}\right)\right|^{2}
\end{array}\right],
\end{equation}
 $g_{\sigma\sigma^{\prime}}=4\pi\hbar^{2}a_{\sigma\sigma^{\prime}}/M$
are the interaction strengths for the intra- ($\sigma=\sigma^{\prime}$)
and inter-species ($\sigma\neq\sigma^{\prime}$), and $a_{\sigma\sigma^{\prime}}$
are corresponding $s$-wave scattering lengths.

In principle, we need to solve a full three-dimensional (3D) problem,
since interatomic interactions couple the motion of atoms in the $x-y$
plane with that along the $z$ axis, although the spin-orbital-angular-momentum (SOAM) coupling is only in the $x-y$ plane. Such a full
simulation would become very time consuming. However, we may assume
that the motion of atoms along the $z$ axis is not affected by the
SOAM coupling in the $x-y$ plane, even in the presence of interatomic
interactions. This means that the wave function may approximately
be written as
\begin{equation}
\Psi\left({\bf r}\right)=\left[\begin{array}{c}
\psi_{\uparrow}^{(2D)}\left(\boldsymbol{\rho}\right)\\
\psi_{\downarrow}^{(2D)}\left(\boldsymbol{\rho}\right)
\end{array}\right]\chi\left(z\right),\label{eq:A10}
\end{equation}
 where $\boldsymbol{\rho}=\left(\rho,\phi\right)$, $\chi\left(z\right)$
takes the form of that in the absence of SOAM coupling \cite{Pethick2008BSpp},
i.e.,
\begin{equation}
\chi\left(z\right)=\frac{e^{-z^{2}/2b^{2}}}{\pi^{1/4}b^{1/2}},
\end{equation}
 and $b$ is the width of the atomic cloud along the $z$ axis before
switching on the SOAM coupling. For our $^{87}$Rb experiment, in
which the total atom number is $N\approx10^{5}$, $a_{\downarrow\downarrow}=a_{\uparrow\downarrow}=100.4a_{B}$,
$a_{\uparrow\uparrow}=100.86a_{B}$, and $a_{B}$ is Bohr's radius,
we find that $b/d\approx3.22622$, and $d=\sqrt{\hbar/M\omega}$ is
the harmonic length. Such a treatment might be a good approximation
when interatomic interactions are weak.

Inserting the ansatz (\ref{eq:A10}) into GP equation, and integrating
both sides with respect to $z$, we obtain an effective two-dimensional
(2D) GP equation, i.e.,
\begin{equation}
\hat{H}_{2D}\left[\begin{array}{c}
\psi_{\uparrow}^{(2D)}\left(\boldsymbol{\rho}\right)\\
\psi_{\downarrow}^{(2D)}\left(\boldsymbol{\rho}\right)
\end{array}\right]=\mu_{2D}\left[\begin{array}{c}
\psi_{\uparrow}^{(2D)}\left(\boldsymbol{\rho}\right)\\
\psi_{\downarrow}^{(2D)}\left(\boldsymbol{\rho}\right)
\end{array}\right],
\end{equation}
 where the reduced 2D Hamiltonian takes the form
\begin{equation}
\hat{H}_{2D}=\hat{H}_{\boldsymbol{\rho}}+\frac{1}{\sqrt{2\pi}b}\left[\begin{array}{cc}
g_{\uparrow\uparrow}\left|\psi_{\uparrow}\left(\boldsymbol{\rho}\right)\right|^{2}+g_{\uparrow\downarrow}\left|\psi_{\downarrow}\left(\boldsymbol{\rho}\right)\right|^{2} & 0\\
0 & g_{\downarrow\downarrow}\left|\psi_{\downarrow}\left(\boldsymbol{\rho}\right)\right|^{2}+g_{\uparrow\downarrow}\left|\psi_{\uparrow}\left(\boldsymbol{\rho}\right)\right|^{2}
\end{array}\right],
\end{equation}
\begin{equation}
\hat{H}_{\boldsymbol{\rho}}=-\frac{\hbar^{2}}{2M}\nabla_{\boldsymbol{\rho}}^{2}+\frac{1}{2}M\omega^{2}\rho^{2}+\frac{\delta}{2}\hat{\sigma}_{z}+\Omega\left(\rho\right)\hat{\sigma}_{x}-\frac{l\hbar}{M\rho^{2}}\hat{L}_{z}\hat{\sigma}_{z}+\frac{\left(l\hbar\right)^{2}}{2M\rho^{2}},
\end{equation}
 and
\begin{equation}
\mu_{2D}=\mu-\frac{1}{4}\left(\frac{d^{2}}{b^{2}}+\frac{b^{2}}{d^{2}}\right)\hbar\omega.
\end{equation}

In the presence of interatomic interactions, it has been shown that
different ground-state quantum phases can still been distinguished
by quasi-orbital angular momentum $l_{z}$ except the stripe phase
\cite{Pu2015PRAsoamSpp,Zhang2015PRAsoamphaseSpp,Pu2016PRAsoamSpp}. Leaving the stripe phase untouched
in this work, let us consider the ground-state quantum phases with
definite quasi-orbital angular momentum $l_{z}$. In this case, the
wave function $\Psi\left({\bf r}\right)$ can be written as
\begin{equation}
\Psi\left({\bf r}\right)=\left[\begin{array}{c}
\varphi_{\uparrow}\left(\rho\right)\\
\varphi_{\downarrow}\left(\rho\right)
\end{array}\right]\frac{e^{il_{z}\phi}}{\sqrt{2\pi}}\chi\left(z\right),
\end{equation}
 then the GP equation can further be reduced to a simple one-dimensional
(1D) equation related to the radial wave function $\varphi_{\sigma}\left(\rho\right)$.
Subsequently, we can easily solve such 1D radial equation simply by
using the finite-difference method, and obtain all the eigen spectrum
and wave functions with respect to different $l_{z}$. Then the ground
state of the system can easily be identified as well as the corresponding
$l_{z}$. With the ground-state wave function in hands, the spin polarization
can easily be obtained,
\begin{equation}
\left\langle \sigma_{z}\right\rangle =\frac{\int d{\bf r}\left[\left|\psi_{\uparrow}\left({\bf r}\right)\right|^{2}-\left|\psi_{\downarrow}\left({\bf r}\right)\right|^{2}\right]}{N},
\end{equation}
 where $N$ is the total particle number, and the wave function is
normalized to $N$, i.e., $\int d{\bf r}\left[\left|\psi_{\uparrow}\left({\bf r}\right)\right|^{2}+\left|\psi_{\downarrow}\left({\bf r}\right)\right|^{2}\right]=N$.

\subsection*{Spin polarization at finite temperature}

Theoretically, for the ground state, we find that the jump of the
spin polarization is very sharp at the boundary of the phase transition. While the finite temperature affects the behavior of the spin polarization. In the following, we consider the finite-temperature
correction to the spin polarization. According to our previous calculations,
the interaction effect is not significant in our experiment. Therefore,
to a preliminary approximation, we treat the system consisting of
condensate and thermal atoms in the absence of interactions.

In our previous calculation, we obtain all the energy levels as well
as corresponding eigen wave functions for a single atom. Different
single-particle states are characterized by the quasi-orbital angular
momentum $l_{z}$, the energy band number $k$ for given $l_{z}$,
and the harmonic quantum number $q$ along the $z$ axis. Then the
single-particle eigen wave function may be written as
\begin{equation}
\Psi_{l_{z}kq}\left({\bf r}\right)=\left[\begin{array}{c}
\varphi_{l_{z}k,\uparrow}\left(\rho\right)\\
\varphi_{l_{z}k,\downarrow}\left(\rho\right)
\end{array}\right]\frac{e^{il_{z}\phi}}{\sqrt{2\pi}}\chi_{q}\left(z\right),
\end{equation}
 where $\varphi_{l_{z}k,\sigma}\left(\rho\right)$ is the radial function
for the spin-$\sigma$ component,
\begin{equation}
\chi_{q}\left(z\right)=\frac{1}{\left(\sqrt{\pi}d\cdot2^{q}q!\right)^{1/2}}e^{-z^{2}/2d^{2}}H_{q}\left(\frac{z}{d}\right)
\end{equation}
is the eigen wave function of 1D harmonic oscillators in the $z$-direction, where $H_{q}\left(\cdot\right)$ is the Hermite polynomial, and $d=\sqrt{\hbar/M\omega}$
is the harmonic length. The corresponding eigen energy is denoted
by $E_{l_{z}kq}$.

For a Bose gas, the Bose-Einstein distribution reads
\begin{equation}
f_{s}\left(E_{s}\right)=\frac{1}{\exp\left[\left(E_{s}-\mu\right)/k_{B}T\right]-1}
\end{equation}
for a single-particle state $s$ with energy $E_{s}$, where $\mu$
is the single-particle chemical potential, $T$ is the temperature,
and $k_{B}$ is the Boltzmann constant.

At finite temperature $0<T<T_{c}$, the chemical potential equals to the ground-state energy,
i.e., $\mu=\varepsilon_0$, and the system consists of the condensate
and thermal atoms. Then the number of thermal atoms is
\begin{equation}
N_{th}=\sum_{\left(l_{z}kq\right)^{\prime}}\left[\exp\left(\frac{E_{l_{z}kq}-\varepsilon_{0}}{k_{B}T}\right)-1\right]^{-1},
\end{equation}
where the summation $\sum_{\left(l_{z}kq\right)^{\prime}}$ is over all the excited single-particle states
except the ground state. Then the atom number of the condensate is
$N_{0}=N-N_{th}$. Therefore, the average value of the spin polarization
should be
\begin{equation}
\left\langle \sigma_{z}\right\rangle =\frac{1}{N}\left[N_{0}\left\langle \sigma_{z}\right\rangle _{0}+\sum_{\left(l_{z}kq\right)^{\prime}}\left\langle \sigma_{z}\right\rangle _{l_{z}kq}\left[\exp\left(\frac{E_{l_{z}kq}-\varepsilon_{0}}{k_{B}T}\right)-1\right]^{-1}\right],
\end{equation}
 where $\left\langle \sigma_{z}\right\rangle _{0}$ and $\left\langle \sigma_{z}\right\rangle _{l_{z}kq}$
are the expected values of the spin polarization of the condensate
and of the excited state $\Psi_{l_{z}kq}\left({\bf r}\right)$, respectively.

\end{widetext}

\end{document}